\documentclass[	DIV=calc,%
							paper=a4,%
							fontsize=11pt,%
							twocolumn]{scrartcl}	 					

\usepackage{lipsum}													

\usepackage[english]{babel}										
\usepackage[protrusion=true,expansion=true]{microtype}				
\usepackage{amsmath,amsfonts,amsthm}					
\usepackage[pdftex]{graphicx}									
\usepackage[svgnames]{xcolor}									
\usepackage[hang, small,labelfont=bf,up,textfont=it,up]{caption}	
\usepackage{epstopdf}												
\usepackage{subfig}													
\usepackage{booktabs}												
\usepackage{fix-cm}													

\usepackage{sectsty}													
\allsectionsfont{
	\usefont{OT1}{phv}{b}{n}
	}

\sectionfont{
	\usefont{OT1}{phv}{b}{n}
	}

\usepackage{fancyhdr}												
\usepackage{lastpage}	

\usepackage{lettrine}
\newcommand{\initial}[1]{%
     \lettrine[lines=3,lhang=0.3,nindent=0em]{
     				\color{DarkGoldenrod}
     				{\textsf{#1}}}{}}

\usepackage{titling}															

\newcommand{\HorRule}{\color{DarkGoldenrod}
									  	\rule{\linewidth}{1pt}%
										}
\pretitle{\vspace{-30pt} \begin{flushleft} \HorRule 
				\fontsize{50}{50} \usefont{OT1}{phv}{b}{n} \color{DarkRed} \selectfont 
				}
\title{Review: Noise and artifact reduction for MRI using deep learning}					
\posttitle{\par\end{flushleft}\vskip 0.5em}

\preauthor{\begin{flushleft}
					\large \lineskip 0.5em \usefont{OT1}{phv}{b}{sl} \color{DarkRed}}
\author{Daiki Tamada, }											
\postauthor{\footnotesize \usefont{OT1}{phv}{m}{sl} \color{Black} 
					Department of Radiology, University of Yamanashi, Chuo, Yamanashi, 409-3898, Japan 								
					\par\end{flushleft}\HorRule}
\date{Submitted to  Magnetic Resonance in Medical Sciences on 2/27/2020}																			

\begin{document}
\maketitle
\thispagestyle{fancy} 			
\initial{F}\textbf{or several years, numerous attempts have been made to reduce noise and artifacts in MRI. Although there have been many successful methods to address these problems, practical implementation for clinical images is still challenging because of its complicated mechanism. Recently, deep learning received considerable attention, emerging as a machine learning approach in delivering robust MR image processing. The purpose here is therefore to explore further and review noise and artifact reduction using deep learning for MRI.}

\section*{Introduction}
MRI suffers from various kinds of noise and artifacts because of the nature of the NMR signal detection and spatial encoding scheme\textsuperscript{1}. There are several reasons that MRI is sensitive to these undesirable errors compared to other modalities. For instance, hardware-induced errors often stem from a complicated acquisition scheme that depends on a reliable spectrometer, RF coils, and magnetic fields. In addition, a long acquisition time is also an essential reason. Body motion, including respiratory and cardiac motion and B0 drift during the scan, can degrade the image quality as well. Notwithstanding the intense dedication of MR vendors to improve these fundamental problems, noise and artifact are still non-negligible degradations to MR images.

Generally, thermal noise is derived from the human body and MRI system itself, such as RF coils, transmission lines, receiver circuits, and so on. Moreover, external RF noise can be a source of the noise. To make matters worse, some sequences such as diffusion-weighted imaging suffer from low signal-to-noise ratio (SNR) because of their weak signal intensity. To improve the SNR of MR images, the signal averaging approach is widely used, although it requires a longer scan time as the number of excitation (NEX) increases. To address these hitches, many advanced filtering methods have been developed using the underlying assumption of MR images such as noise distribution and structures \textsuperscript{2}. However, the violation of the assumption often induces the undesirable change of texture. In addition to this, some algorithms utilize iterative reconstruction, which yield long computation time.

Artifacts in MRI are caused by various reasons such as external errors and inappropriate spatial encoding. For example, it is widely known that undersampling in Cartesian imaging results in aliasing artifacts. In this case, compressed sensing (CS) reconstruction is generally used to remove such artifacts. Streak artifact is caused by undersampling, motion, and temporal intensity change owing to contrast agent in radial sampling. Advanced imaging techniques, combined with golden angle trajectory and CS reconstruction, have shown excellent performance against these problems\textsuperscript{3-7}. Lack of high-frequency signal leads to Gibbs artifact, which is also called truncation, or ringing artifact. Simple k-space domain filtering is often used although it may cause blurring. Body imaging suffers from motion artifacts that arise from respiratory and cardiac motion. Respiratory triggering\textsuperscript{8}, fast acquisition\textsuperscript{9}, and radial sampling\textsuperscript{3} are clinically available to suppress artifacts.

Further, retrospective correction\textsuperscript{10} approaches have been proposed, although it is still in the research stage. Inhomogeneous B0 field or off-resonance effect results in banding artifacts, which are often seen in balanced steady-state free precession.  The development of robust and practical methods is still challenging despite the aforementioned efforts to remove artifacts.\  

Recently, a deep learning approach, which enables feature extraction and complicated nonlinear image processing, is gaining traction to reduce noise and artifacts in MRI\textsuperscript{11}. Deep learning (DL) is a successful machine learning technique based on the neural network used for segmentation, lesion detection, and reconstruction for MRI. In this study, we review the deep learning techniques and applications for reducing noise and artifacts in MRI. First, we discuss the deep learning architecture used for noise and artifact reduction. The following section describes state-of-the-art applications.

\begin{figure*}[bt]
\centering
\includegraphics[width=16cm]{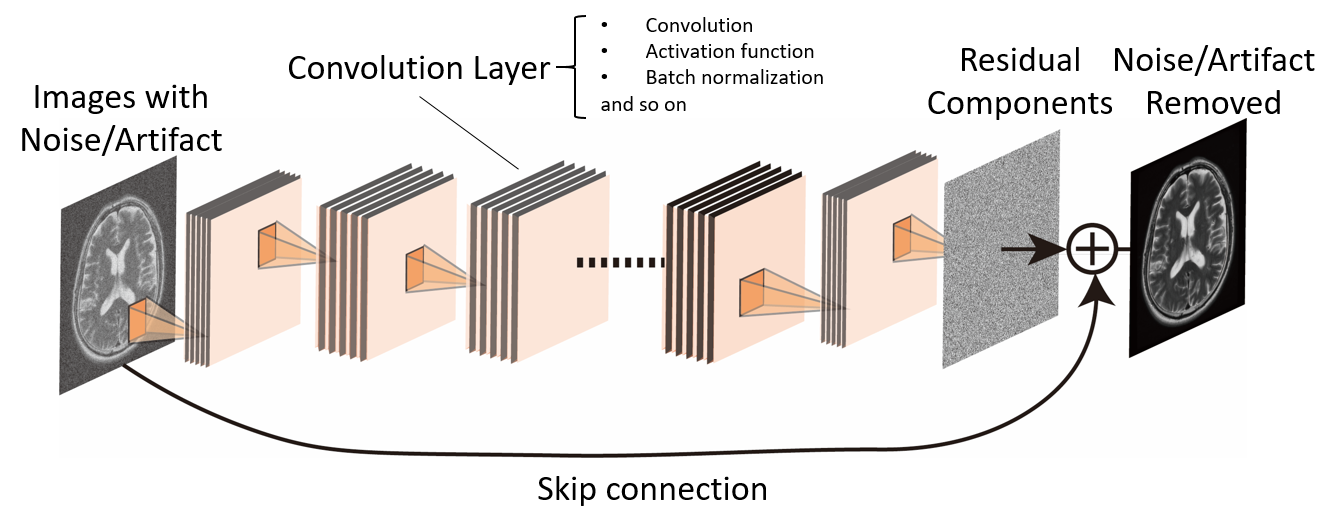}
\caption{Example of a single-scale CNN structure consisting of multiple convolution layers. Each convolution layer has a set of filters, which is followed by activation function such as ReLU, to extract features of input images. The residual learning approach is usually adopted for noise and artifact reduction. Noise or artifact reduced images are calculated by subtracting the output of the network from input images.}
\label{Fig1}
\end{figure*}

\section{Deep Learning Architecture}

Many DL networks have been proposed for noise and artifact reduction. Convolutional neural networks (CNN), which has a small receptive field, is widely used for noise reduction because Gaussian noise is incoherent and position-independent. However, only the networks with a larger receptive field are often adopted to improve artifacts because most of the artifacts are distributed globally. In this section, we provide an overview of deep learning architecture commonly used for noise and artifact reduction.

Single-scale CNN (SCNN), which is widely used in the various fields shown in Fig. 1, is efficient in improving noisy images\textsuperscript{12, 13}. SCNN consists of more than one convolution layer without a down- or up-sampling structure, such as pooling and stride. In most cases, residual components such as noise or artifact images are used as the output because it is well-known that the manifold of the residual components has a simpler structure compared to that of the clean image components\textsuperscript{14}, resulting in faster conversion and better generalization of the network. Denoising convolutional neural networks (DnCNN) is also a thriving network used for image denoising through a process of learning residual components of images, which is noise. Besides these, batch normalization is used to improve training time for the network. ResNet is commonly used to achieve a complex and deeper structure of the network to overcome the vanishing gradient problem. ResNet is also used for artifact reduction because it has a large receptive field that covers whole input images owing to a large number of layers \textsuperscript{15}.

Autoencoders can extract clean features from noisy images, similar to the principal component analysis procedure, which is used to extract significant information from high-dimensional datasets \textsuperscript{16}. Given that Gondara first developed an autoencoder-based denoising filter for X-ray images\textsuperscript{17}, many studies have been proposed for CT, MRI, and others, recently\textsuperscript{18, 19}. Autoencoder, which has a different mechanism compared to SCNN, compresses the entire image information into a low-dimensional structure to effectively learn underlying manifold. Unfortunately, autoencoders may lose important information such as edge or fine structure, if the input image is not a redundant representation.

U-net, often used for image segmentation, has a large receptive field by utilizing multi-scale features\textsuperscript{20}. This network has asymmetric up- and down-sampling structures with a skip connection, which concatenates each layer, as shown in Fig.  2. The skip connection contributes to capturing localized features such as the fine structure of images. For the artifact reduction, U-net is the most popular network because the large receptive field is generally required to extract the artifact that is commonly observed globally. The study by Lee et al. implied that U-net gave a better de-aliasing performance as compared to SCNN\textsuperscript{14}, although a relatively large number of trainable parameters is required for U-net. 

\begin{figure*}[bt]
\centering
\includegraphics[width=16cm]{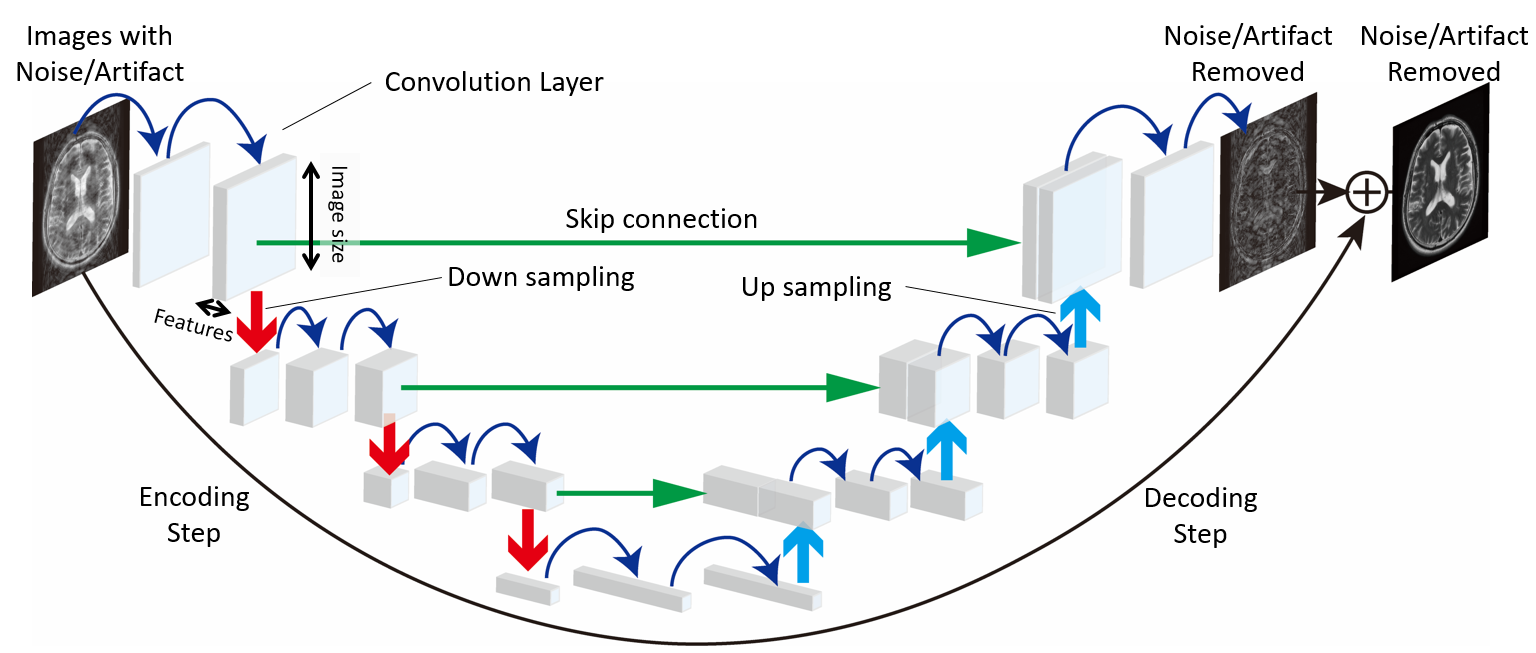}
\caption{Overview of U-net, which has up- and down-sampling structure - the same as autoencoder - used for noise and artifact reduction. Skip connections by copying features from earlier layers into later layers are used to preserve high-frequency information, which is often lost in the down-sampling step. This network is commonly used for removing artifact reduction, and requires a larger receptive field compared to noise reduction.}
\label{Fig2}
\end{figure*}

Generative adversarial networks (GAN) approach is a promising learning technique for removing noise and artifacts\textsuperscript{21}. GAN consists of two separate networks: generator and discriminator. Generator is the network that creates the output images, noise- or artifact-free images, whereas discriminator acts as a classifier to determine whether the generated output is real or fake. Feedback from the discriminator contributes to updating the network of the generator to create convincing fake images to fool the discriminator. U-net is widely used for the generator.

\section{Applications}

\begin{figure*}[bt]
\centering
\includegraphics[width=16cm]{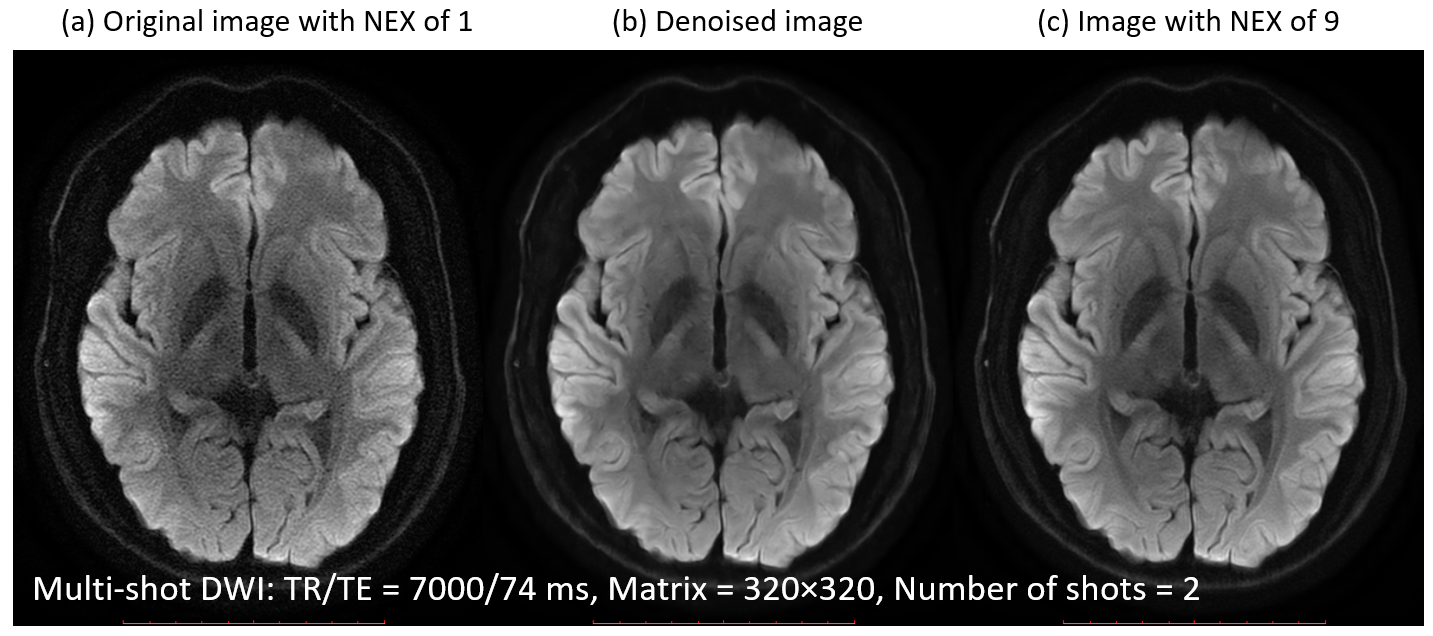}
\caption{Specimen images of denoising brain DWI using DnCNN, the Noise2Noise approach. This network includes 17 convolution layers with batch normalization and ReLU. Multi-shot DWI images (a) acquired with NEX of 1 were denoised using the network. PSNR analysis between the denoised and high-NEX images indicated that the network successfully reduced the noise component.}
\label{Fig3}
\end{figure*}

Many applications have been proposed using the above successful DL networks for noise and artifact reduction. Although most parts of them are still research stage, promising results were demonstrated with a limited size of datasets. In this section, we introduce state-of-art applications shown in Tab. 1.

Some DL networks focus on denoising for brain MR imaging. Bermudez et al. proposed an autoencoder with skip connections for T1-weighted (T1w) imaging of the brain\textsuperscript{19}. The developed network, trained with 528 T1w images, significantly improved the image quality based on PSNR analysis. DnCNN is also widely used for denoising of brain MR images\textsuperscript{22, 23}. Denoising using DL is used not only for the improvement of image quality but also for reducing scan time. Currently, MR images with higher resolutions are required to achieve better diagnostic performance. However, improvement of patient experience, particularly scan time, in the MR scanner is also an urgent priority. To overcome these conflicting problems, many DL-based denoising approaches have been developed. Kawamura et al. proposed the DnCNN-based approach for denoising multi-shot diffusion-weighted imaging for the brain. The study showed that highly swift DWI could be achieved by reducing the number of excitations (NEX), as shown in Fig. 3.  Kidoh et al. proposed the SCNN network, which can be used for multiple sequences such as T1w, T2w, FLAIR, and MPRAGE\textsuperscript{24}. The proposed network utilizes a discrete cosine transform (DCT) layer to separate zero and high-frequency components. The separated high-frequency components go through 22 convolution layers, whereas the zero-frequency component is connected to the last layer. This structure may contribute to robust denoising performance because it doesn’t use contrast information of images. Imaging can be achieved at a pace of more than twofold by using the proposed method.

Moreover, Arterial spin labeling (ASL) perfusion imaging, which generally suffers from low SNR, with DL denoising is also a promising application\textsuperscript{25, 26}. Xie et al. developed the DL denoising approach for ASL using a wide interface network (WIN)\textsuperscript{25}. The WIN is a residual learning CNN with a large size of convolution kernel and filters.  Quantitative analysis based on SNR suggested that the proposed method can reduce the scan time for ASL by reducing NEX. Kim et al. proposed a network consisting of two separate pathways for extracting multi-scale features\textsuperscript{26}. The pathway with convolution kernel, batch normalization, and ReLU is used for the local structure of images. Meanwhile, the other pathway uses a dilated pathway to increase the receptive field in order to extract large-scale information. Evaluation by a radiologist implied that the ASL images with low NEX were significantly improved by using this proposed CNN method.

Further, Streak artifact reduction is one of the most successful applications using DL\textsuperscript{27}. There are many studies for reducing streak artifact for computed tomography (CT) images because streak artifact on CT is a very common undesirable image degradation\textsuperscript{28-30}. Recently, radial sampling for abdominal MR imaging is getting attention because of its robustness against respiratory induced motion\textsuperscript{3, 5, 6}. In the case of radial sampling, streak artifact can be observed in MR images as well as CT. Han et al. developed U-net to remove streak artifacts for the brain and abdominal MR images; a network adopted residual learning with pre-training using CT and MRI datasets. This study indicated that the DL-based method could be superior to conventional compressed sensing algorithms. 

Several studies of retrospective and blinded motion artifact correction techniques for the brain\textsuperscript{31-33}, c-spine\textsuperscript{34}, liver\textsuperscript{35, 36}, upper abdomen\textsuperscript{37}, and cardiac\textsuperscript{38} imaging  have been proposed.  Pawar et al. proposed the motion artifact reduction technique using the U-net for MPRAGE images of the brain. Evaluation based on naturalness image quality evaluator (NIQE)\textsuperscript{39}, which is reference-free image quality metrics, revealed the proposed method improved the motion artifact. Tamada et al. also proposed a motion artifact reduction for DCE-MRI of the liver using DnCNN-based network with multi-channel input and output. A training dataset was generated by simulating the respiratory motion. The results indicated the DL approach possessed the ability to remove artifacts, as show in Fig. 4. It is a challenge to acquire artifact-free images under the influence of cardiac motion.  Oksuz et al. proposed the reconstruction technique for cardiac imaging by combining the 3D CNN and the recurrent convolutional neural network (RCNN)\textsuperscript{40} used for 2D spatiotemporal imaging. The 3D CNN is adopted to detect the corrupted k-space lines by the motion whereas RCNN removes the motion artifact coming from the corrupted lines. 

Besides, some applications are used for removing aliasing artifact alternatives to CS reconstruction\textsuperscript{41-43}. CS reconstruction, which is a fast MRI technique using random undersampling and redundancy of MR images, generally requires long computation time. DL approaches can contribute to the reduction of computation cost for the undersampled datasets. A de-aliasing method using the U-net for undersampled k-space was proposed to achieve efficient reconstruction by Hyun et al. \textsuperscript{41}. Yang et al. proposed a GAN-based reconstruction method for undersampled images\textsuperscript{42}. Recently, a model-based DL network, which has unrolled architecture, was proposed to reduce trainable parameters of the network\textsuperscript{43}.

The DL-based approach is also useful in solving other complicated problems such as removing Gibbs\textsuperscript{44-47}, metal\textsuperscript{48-50}, and banding artifact\textsuperscript{51}. Zhang et al. proposed the CNN-based Gibbs artifact, which can be an alternative to conventional k-space domain filtering methods\textsuperscript{45}. The results indicated that the proposed method removed artifacts without additional blurring, which is often observed in traditional methods. Kwon proposed metal artifact reduction using the U-net and bipolar readout imaging to improve blurring and distortion caused by a strong off-resonance effect\textsuperscript{48}. Seo et al. developed the CNN to correct metal artifact of MR images acquired using slice encoding for metal artifact correction\textsuperscript{49}. 

\begin{figure*}[bt]
\centering
\includegraphics[width=16cm]{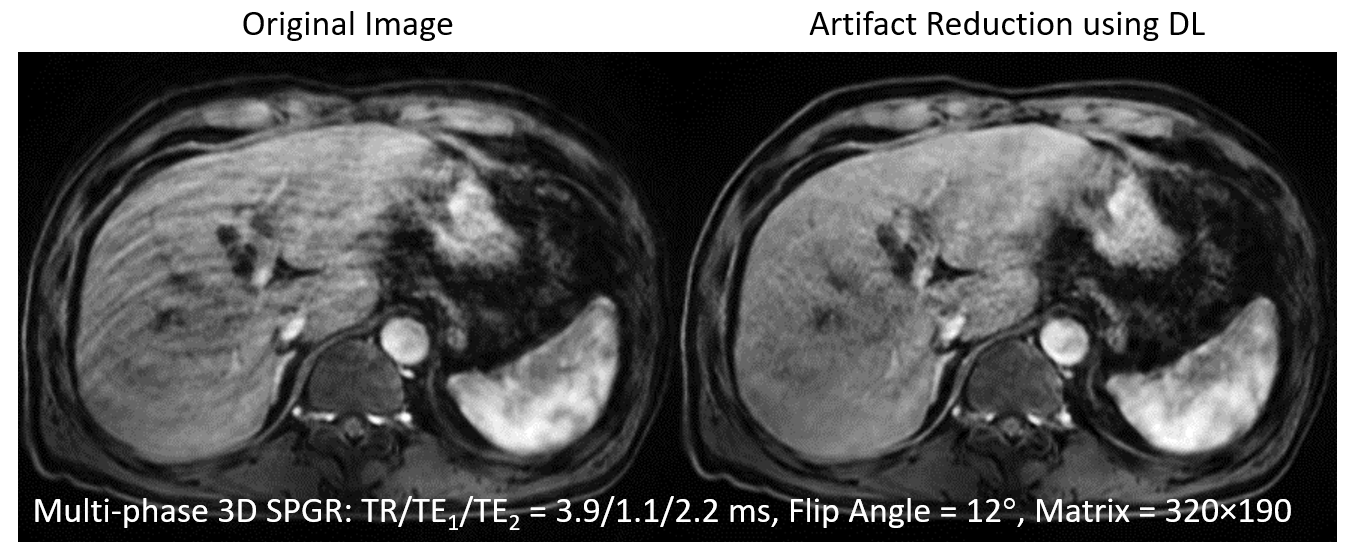}
\caption{Sample depictions of aliasing reduction of the liver using multi-channel DnCNN, which have seven layers of convolution layers. Multi-contrast images obtained from DCE-MRI were used for the input of the network. MR images with respiratory artifact and blurring were significantly improved by via this network. The acquisition was implemented using the multi-phase 3D T1-weighted spoiled gradient-echo sequence with a dual-echo bipolar readout.}
\label{Fig4}
\end{figure*}

\section{Discussion}

In this paper, we summarized DL networks and its applications used for noise and artifact reduction. Typical networks including SCNN, U-net, autoencoder, GAN architectures, which have characteristics of structure different from each other, were explained. Further, we reviewed applications using these networks. The studies reviewed here demonstrate the excellent performance of noise and artifact reduction for MRI. These results revealed that DL-base approaches can be used to remove complicated artifacts such as aliasing, streaking, and so on, which are still challenging problems using conventional methods.

Although many promising techniques have been proposed, challenges of removing noise and artifact using DL exist. To begin with, it is challenging to prepare training datasets for some cases. For noise reduction, images with a large number of NEX are often used as clean datasets. However, motions, such as respiratory, cardiac, and CSF pulsation, during a scan can lead to blurring and undesirable image degradation. Noise2Noise approach could be one of the practical solutions for noise reduction\textsuperscript{52}. Nonetheless, datasets generated by simulation are generally used for artifact reduction. In this case, the discrepancy between the simulated and acquired datasets leads to poor filtering performance. 

Furthermore, it is difficult to conduct appropriate image quality evaluation for noise and artifact reduction. MSE, PSNR, and SSIM are widely used to evaluate the image quality, particularly for noise reduction networks. Conversely, there is difficulty in using these conventional metrics for artifact reduction because it is challenging to prepare ground truth datasets that have no artifact images. To evaluate the image quality without ground truth, reference-free metrics such as Blind/Referenceless Image Spatial Quality Evaluator\textsuperscript{53} and NIQE\textsuperscript{39} are a reasonable alternative to conventional metrics. However, clinical evaluation is essential for practical use because these metrics offer excellent performance for natural images. It is essential to pay attention to the contrast, texture, and fine structure after DL-based filtering because inappropriate training, such as the use of a small number of datasets and wrong noise or artifact modeling, could lead to a critical change of contexture of images.

\section{Conclusion and Outlook}

In this study, we reviewed DL-based methods to reduce noise and artifact for MR images. Because of the recent advancement of deep learning technology, a wide variety of techniques are proposed to serve these purposes. However, even though DL-based methods are getting popular in the MR community owing to its excellent performance and low computational cost, there are still issues to be solved for practical use. Nevertheless, some applications are overcoming these setbacks and are about to come onto the clinical site as products from MR vendors. The DL-based method will therefore be widely used in a range of MR applications shortly.

\section{References}

\noindent \textbf{1.}\hspace{0.25cm} Bellon EM, Haacke EM, Coleman PE, Sacco DC, Steiger DA, Gangarosa RE. MR artifacts: a review. American Journal of Roentgenology 1986; 147:1271-1281.

\noindent \textbf{2.}\hspace{0.25cm} Balafar M. Review of noise reducing algorithms for brain MRI images. methods 2012; 10:11.

\noindent \textbf{3.}\hspace{0.25cm} Chandarana H, Feng L, Block TK, et al. Free-breathing contrast-enhanced multiphase MRI of the liver using a combination of compressed sensing, parallel imaging, and golden-angle radial sampling 2013; 48.

\noindent \textbf{4.}\hspace{0.25cm} Chandarana H, Block KT, Winfeld MJ, et al. Free-breathing contrast-enhanced T1-weighted gradient-echo imaging with radial k-space sampling for paediatric abdominopelvic MRI 2014; 24:320-326.

\noindent \textbf{5.}\hspace{0.25cm} Chandarana H, Block TK, Rosenkrantz AB, et al. Free-breathing radial 3D fat-suppressed T1-weighted gradient echo sequence: a viable alternative for contrast-enhanced liver imaging in patients unable to suspend respiration 2011; 46:648-653.

\noindent \textbf{6.}\hspace{0.25cm} Feng L, Grimm R, Block KT, et al. Golden‐angle radial sparse parallel MRI: combination of compressed sensing, parallel imaging, and golden‐angle radial sampling for fast and flexible dynamic volumetric MRI 2014; 72:707-717.

\noindent \textbf{7.}\hspace{0.25cm} Feng L, Axel L, Chandarana H, Block KT, Sodickson DK, Otazo RJMrim. XD‐GRASP: golden‐angle radial MRI with reconstruction of extra motion‐state dimensions using compressed sensing 2016; 75:775-788.

\noindent \textbf{8.}\hspace{0.25cm} Chavhan GB, Babyn PS, Vasanawala SSJR. Abdominal MR imaging in children: motion compensation, sequence optimization, and protocol organization 2013; 33:703-719.

\noindent \textbf{9.}\hspace{0.25cm} Zhang T, Chowdhury S, Lustig M, et al. Clinical performance of contrast enhanced abdominal pediatric MRI with fast combined parallel imaging compressed sensing reconstruction 2014; 40:13-25.

\noindent \textbf{10.}\hspace{0.25cm} Cheng JY, Alley MT, Cunningham CH, Vasanawala SS, Pauly JM, Lustig MJMrim. Nonrigid motion correction in 3D using autofocusing withlocalized linear translations 2012; 68:1785-1797.

\noindent \textbf{11.}\hspace{0.25cm} Lundervold AS, Lundervold A. An overview of deep learning in medical imaging focusing on MRI. Zeitschrift für Medizinische Physik 2019; 29:102-127.

\noindent \textbf{12.}\hspace{0.25cm} Kim J, Kwon Lee J, Mu Lee K. Accurate image super-resolution using very deep convolutional networks. Proceedings of the IEEE conference on computer vision and pattern recognition, 2016; 1646-1654.

\noindent \textbf{13.}\hspace{0.25cm} Zhang K, Zuo W, Chen Y, Meng D, Zhang LJIToIP. Beyond a gaussian denoiser: Residual learning of deep cnn for image denoising 2017; 26:3142-3155.

\noindent \textbf{14.}\hspace{0.25cm} Lee D, Yoo J, Ye JC. Deep residual learning for compressed sensing MRI. Biomedical Imaging (ISBI 2017), 2017 IEEE 14th International Symposium on, 2017; 15-18.

\noindent \textbf{15.}\hspace{0.25cm} Ghodrati V, Shao J, Bydder M, et al. MR image reconstruction using deep learning: evaluation of network structure and loss functions. Quantitative imaging in medicine and surgery 2019; 9:1516.

\noindent \textbf{16.}\hspace{0.25cm} Vincent P, Larochelle H, Bengio Y, Manzagol P-A. Extracting and composing robust features with denoising autoencoders. Proceedings of the 25th international conference on Machine learning, 2008; 1096-1103.

\noindent \textbf{17.}\hspace{0.25cm} Gondara L. Medical image denoising using convolutional denoising autoencoders. 2016 IEEE 16th International Conference on Data Mining Workshops (ICDMW), 2016; 241-246.

\noindent \textbf{18.}\hspace{0.25cm} Nishio M, Nagashima C, Hirabayashi S, et al. Convolutional auto-encoder for image denoising of ultra-low-dose CT. Heliyon 2017; 3:e00393.

\noindent \textbf{19.}\hspace{0.25cm} Bermudez C, Plassard AJ, Davis LT, Newton AT, Resnick SM, Landman BA. Learning implicit brain MRI manifolds with deep learning. Medical Imaging 2018: Image Processing, 2018; 105741L.

\noindent \textbf{20.}\hspace{0.25cm} Ronneberger O, Fischer P, Brox T. U-net: Convolutional networks for biomedical image segmentation. International Conference on Medical image computing and computer-assisted intervention, 2015; 234-241.

\noindent \textbf{21.}\hspace{0.25cm} Goodfellow I, Pouget-Abadie J, Mirza M, et al. Generative adversarial nets. Advances in neural information processing systems, 2014; 2672-2680.

\noindent \textbf{22.}\hspace{0.25cm} Jiang D, Dou W, Vosters L, Xu X, Sun Y, Tan T. Denoising of 3D magnetic resonance images with multi-channel residual learning of convolutional neural network. Japanese journal of radiology 2018; 36:566-574.

\noindent \textbf{23.}\hspace{0.25cm} Rao GS, Srinivas B. De-noising of MRI Brain Tumor image using Deep Convolutional Neural Network. Available at SSRN 3357284 2019.

\noindent \textbf{24.}\hspace{0.25cm} Kidoh M, Shinoda K, Kitajima M, et al. Deep Learning Based Noise Reduction for Brain MR Imaging: Tests on Phantoms and Healthy Volunteers. Magnetic Resonance in Medical Sciences 2019:mp. 2019-0018.

\noindent \textbf{25.}\hspace{0.25cm} Xie D, Bai L, Wang Z. Denoising arterial spin labeling cerebral blood flow images using deep learning. arXiv preprint arXiv:180109672 2018.

\noindent \textbf{26.}\hspace{0.25cm} Kim KH, Choi SH, Park S-H. Improving arterial spin labeling by using deep learning. Radiology 2018; 287:658-666.

\noindent \textbf{27.}\hspace{0.25cm} Han Y, Yoo J, Kim HH, Shin HJ, Sung K, Ye JC. Deep learning with domain adaptation for accelerated projection‐reconstruction MR. Magnetic resonance in medicine 2018; 80:1189-1205.

\noindent \textbf{28.}\hspace{0.25cm} Gjesteby L, Yang Q, Xi Y, et al. Reducing metal streak artifacts in CT images via deep learning: Pilot results. The 14th international meeting on fully three-dimensional image reconstruction in radiology and nuclear medicine, 2017; 611-614.

\noindent \textbf{29.}\hspace{0.25cm} Han YS, Yoo J, Ye JC. Deep residual learning for compressed sensing CT reconstruction via persistent homology analysis. arXiv preprint arXiv:161106391 2016.

\noindent \textbf{30.}\hspace{0.25cm} Han Y, Ye JC. Framing U-Net via deep convolutional framelets: Application to sparse-view CT. Ieee T Med Imaging 2018; 37:1418-1429.

\noindent \textbf{31.}\hspace{0.25cm} Pawar K, Chen Z, Shah NJ, Egan GF. Moconet: Motion correction in 3D MPRAGE images using a convolutional neural network approach. arXiv preprint arXiv:180710831 2018.

\noindent \textbf{32.}\hspace{0.25cm} Duffy BA, Zhang W, Tang H, et al. Retrospective correction of motion artifact affected structural MRI images using deep learning of simulated motion 2018.

\noindent \textbf{33.}\hspace{0.25cm} Johnson PM, Drangova M. Motion correction in MRI using deep learning. ISMRM Scientific Meeting $\&$  Exhibition, Honolulu, 2018; 4098.

\noindent \textbf{34.}\hspace{0.25cm} Lee H, Ryu K, Nam Y, Lee J, Kim D-H. Reduction of respiratory motion artifact in c-spine imaging using deep learning: Is substitution of navigator possible? ISMRM Scientific Meeting $\&$  Exhibition, 2018; 2660.

\noindent \textbf{35.}\hspace{0.25cm} Tamada D, Kromrey M-L, Ichikawa S, Onishi H, Motosugi U. Motion Artifact Reduction Using a Convolutional Neural Network for Dynamic Contrast Enhanced MR Imaging of the Liver. Magnetic Resonance in Medical Sciences 2020; 19:64-76.

\noindent \textbf{36.}\hspace{0.25cm} Tamada D, Onishi H, Motosugi U. Motion Artifact Reduction in Abdominal MR Imaging using the U-NET Network. ICMRM and Scientific Meeting of KSMRM, Seoul, Korea, 2018; PP03–11.

\noindent \textbf{37.}\hspace{0.25cm} Jiang W, Liu Z, Lee K-H, et al. Respiratory motion correction in abdominal MRI using a densely connected U-Net with GAN-guided training. arXiv preprint arXiv:190609745 2019.

\noindent \textbf{38.}\hspace{0.25cm} Oksuz I, Clough J, Ruijsink B, et al. Detection and Correction of Cardiac MRI Motion Artefacts During Reconstruction from k-space. International Conference on Medical Image Computing and Computer-Assisted Intervention, 2019; 695-703.

\noindent \textbf{39.}\hspace{0.25cm} Mittal A, Soundararajan R, Bovik AC. Making a $``$completely blind$"$  image quality analyzer. IEEE Signal Processing Letters 2012; 20:209-212.

\noindent \textbf{40.}\hspace{0.25cm} Qin C, Schlemper J, Caballero J, Price AN, Hajnal JV, Rueckert D. Convolutional recurrent neural networks for dynamic MR image reconstruction. Ieee T Med Imaging 2018; 38:280-290.

\noindent \textbf{41.}\hspace{0.25cm} Hyun CM, Kim HP, Lee SM, Lee S, Seo JKJPim, biology. Deep learning for undersampled MRI reconstruction 2018.

\noindent \textbf{42.}\hspace{0.25cm} Yang G, Yu S, Dong H, et al. DAGAN: Deep de-aliasing generative adversarial networks for fast compressed sensing MRI reconstruction. Ieee T Med Imaging 2017; 37:1310-1321.

\noindent \textbf{43.}\hspace{0.25cm} Aggarwal HK, Mani MP, Jacob M. MoDL: Model-based deep learning architecture for inverse problems. Ieee T Med Imaging 2018; 38:394-405.

\noindent \textbf{44.}\hspace{0.25cm} Zhang Q, Ruan G, Yang W, Zhao K, Wu EX, Feng Y. Gibbs-Ringing Artifact Reduction in MRI via Machine Learning Using Convolutional Neural Network. ISMRM 2018, 2018; 0429.

\noindent \textbf{45.}\hspace{0.25cm} Zhang Q, Ruan G, Yang W, et al. MRI Gibbs‐ringing artifact reduction by means of machine learning using convolutional neural networks. Magnetic resonance in medicine 2019; 82:2133-2145.

\noindent \textbf{46.}\hspace{0.25cm} Muckley MJ, Ades-Aron B, Papaioannou A, et al. Training a Neural Network for Gibbs and Noise Removal in Diffusion MRI. arXiv preprint arXiv:190504176 2019.

\noindent \textbf{47.}\hspace{0.25cm} Zhao X, Zhang H, Zhou Y, Bian W, Zhang T, Zou X. Gibbs-ringing artifact suppression with knowledge transfer from natural images to MR images. Multimedia Tools and Applications 2019:1-23.

\noindent \textbf{48.}\hspace{0.25cm} Kwon K, Kim D, Park H. A Learning-Based Metal Artifacts Correction Method for MRI Using Dual-Polarity Readout Gradients and Simulated Data. International Conference on Medical Image Computing and Computer-Assisted Intervention, 2018; 189-197.

\noindent \textbf{49.}\hspace{0.25cm} Seo S, Do WJ, Luu HM, Kim KH, Choi SH, Park SH. Artificial neural network for Slice Encoding for Metal Artifact Correction (SEMAC) MRI. Magnetic Resonance in Medicine 2019.

\noindent \textbf{50.}\hspace{0.25cm} Kim JW, Kwon K, Kim B, Park H. Attention Guided Metal Artifact Correction in MRI using Deep Neural Networks. arXiv preprint arXiv:191008705 2019.

\noindent \textbf{51.}\hspace{0.25cm} Kim KH, Park S-H. Artificial neural network for suppression of banding artifacts in balanced steady-state free precession MRI. Magnetic resonance imaging 2017; 37:139-146.

\noindent \textbf{52.}\hspace{0.25cm} Lehtinen J, Munkberg J, Hasselgren J, et al. Noise2noise: Learning image restoration without clean data. arXiv preprint arXiv:180304189 2018.

\noindent \textbf{53.}\hspace{0.25cm} Mittal A, Moorthy AK, Bovik AC. No-reference image quality assessment in the spatial domain. IEEE Transactions on image processing 2012; 21:4695-4708.

\begin{table*}[t]
\centering
\caption{Overview of DL studies for noise and artifact reduction}
\label{tab:1}
  \centering
  \begin{tabular}{p{5cm} p{5cm} p{5cm}}
    \toprule
    Purpose & Authors & Network \\
	\midrule
    Denoising for T1-weighted brain images & Camilo Bermudez, et al\textsuperscript{19} & Autoencoder with skip connections \\
    Denoising for brain MRI	& Dongsheng Jiang, et al\textsuperscript{22} & Multi-channel DnCNN\\
    Denoising for ASL &	Ki Hwan Kim et al\textsuperscript{26} & Parallel two CNN with different scale of resolution \\
    Denoising for ASL &	Danfeng Xie, et al\textsuperscript{25} & Wide Inference Network\\
    Denoising for T1-, T2-weighted, and FLAIR brain images &	Masafumi Kidoh, et al\textsuperscript{24}  & Single-scale CNN with DCT\\
    Denoising for multi-shot DWI of the brain &	Motohide Kawamura, et al\textsuperscript{54} & DnCNN with Noise2Noise approach\\
    Streak artifact reduction for radial MRI &	Yoseob Han, et al\textsuperscript{27} &U-net\\
    Motion artifact reduction for brain MRI &	Patricia Johnson, et al\textsuperscript{33} & Single-scale CNN\\
    Motion artifact reduction for DCE-MRI of the liver &	Daiki Tamada, et al\textsuperscript{35} & Multi-channel DnCNN\\
    Motion artifact reduction for abdomen MRI &	Wenhao Jiang, et al\textsuperscript{37} & GAN using U-net as generator network\\
    Motion artifact reduction for cardiac MRI &	Ilkay Oksuz, et al\textsuperscript{38} & 3DCNN and RCNN\\
    Motion artifact reduction for brain MRI	& Ben A Duffy, et al\textsuperscript{32} & GAN using HighRes3dNet as a generator\\
    Motion artifact reduction for cervical spine MRI &	Hongpyo Lee, et al\textsuperscript{34} & U-net\\
    Aliasing artifact reduction for undersampled MRI &	Chang Min Hyun, et al\textsuperscript{41} & U-net\\
    Aliasing artifact reduction for undersampled MRI &	Guang Yang, et al\textsuperscript{42} & GAN using U-net as generator network\\
    Aliasing artifact reduction for undersampled MRI &	Hemant Kumar Aggarwal, et al\textsuperscript{43} & Model-based DL\\
    Gibbs artifact reduction &	Qianqian Zhang, et al\textsuperscript{45} & Single-scale CNN\\
    Gibbs artifact reduction &	Xiaole Zhao, et al\textsuperscript{47} & Enhanced Deep Super-Resolution Network\\
    Metal artifact reduction &	Kinam Kwon, et al\textsuperscript{} & U-net\\
    Metal artifact reduction &	Jee Won Kim, et al\textsuperscript{50} & U-net\\
    Metal artifact reduction for SEMAC & Sunghun Seo, et al\textsuperscript{49} & U-net\\
    Banding artifact reduction for bSSFP &	Ki Hwan Kim, et al\textsuperscript{51} & Multilayer perceptron

  \end{tabular}
\end{table*}

\end{document}